
%
\magnification=\magstep0                                                %
\font\titlefont=cmr10 scaled\magstep3  
\font\titleital=cmmi10 scaled\magstep3             %
\font\greekbold=cmmib10
\newbox\leftpage \newdimen\fullhsize \newdimen\hstitle
\newdimen\hsbody   %
\hoffset=0.0truein \voffset=0.20truein \hsbody=\hsize \hstitle=\hsize
\tolerance=1000\hfuzz=2pt \baselineskip=12pt plus 4pt minus 2pt
%
\global\newcount\secno \global\secno=0 \global\newcount\meqno
\def\newsec#1{\global\advance\secno by1 \xdef\secsym{\ifcase\secno
\or I\or II\or III\or IV\or V\or VI\or VII\or VIII\or IX\or X\fi }%
\xdef\numsecsym{\the\secno}
\global\meqno=1\bigbreak\bigskip\noindent{\bf \secsym. #1}
\par\nobreak\medskip\nobreak}                                             %
\def\appendix#1#2{\xdef\secsym{\hbox{#1}} \xdef\numsecsym{\hbox{#1}}
\global\meqno=1 \bigbreak\bigskip \noindent{\bf Appendix #1. #2}
\par\nobreak\medskip\nobreak}                                             %
\def\eqn#1#2{\xdef
#1{({\rm\numsecsym}.\the\meqno)}\global\advance\meqno by1 %
$$#2\eqno#1$$} %
\global\newcount\ftno \global\ftno=1                                      %
\def\foot#1{{\baselineskip=14pt plus 1pt\footnote{$^{\the\ftno}$}{#1}}
\global\advance\ftno by1}                                                 %
\global\newcount\refno \global\refno=1 \newwrite\rfile                    %
\def\ref#1#2{{[\the\refno]}\nref#1{#2}}
\def\nref#1#2{\xdef#1{\the\refno}
\ifnum\refno=1\immediate\openout\rfile=refs.tmp\fi
\immediate\write\rfile{\noexpand\item{[\the\refno]\ }#2}
\global\advance\refno by1}                                                %

\def\cite#1{[#1]}                                                 %
\def\figures{\centerline{{\bf Figure
Captions}}\medskip\parindent=40pt}
\def\fig#1#2{\medskip\item{Fig.~#1:  }#2}                                 %

%

\def\del{\partial} 

\def\goes{\rightarrow}
\def\vbare{V(\bphi({\bf x}),{\bf x})}
\def\xint{\int\!d^d\!{\bf x} \;}
\def\bx{{\bf x}}
\def\bphi{\hbox{$\phi\textfont1=\greekbold$}}
\def\bk{\hbox{$\kappa\textfont1=\greekbold$}}

\nopagenumbers \hsize=\hsbody \pageno=0 ~ \vfill
\centerline{\titlefont Large $N \textfont1 = \titleital \textfont0 =
\titlefont$ Expansion for $4-\epsilon\textfont1 = \titleital \textfont0 =
\titlefont$ Dimensional}
\centerline{\titlefont Oriented Manifolds in  Random Media}
\vfill \centerline{{\sl Leon Balents and Daniel S. Fisher}}
\centerline{Department of Physics}
\centerline{Harvard University}
\centerline{Cambridge, MA 02138}
\vfill\centerline{\bf ABSTRACT}
\nobreak\medskip\nobreak\par
{
\noindent The equilibrium statistical mechanics of a $d$
dimensional ``oriented'' manifold in an $N+d$ dimensional random
medium are analyzed in $d=4-\epsilon$ dimensions.  For $N=1$, this
problem describes an interface pinned by impurities.  For $d=1$, the
model becomes identical to the directed polymer in a random medium.
Here, we generalize the functional renormalization group method used
previously to study the interface problem, and extract the behavior in
the double limit $\epsilon$ small and $N$ large, finding non-analytic
corrections in $1/N$.  For short-range disorder, the interface width
scales as $w \sim L^\zeta$, with \break $\zeta = {\epsilon
\over {N+4}}\left\{1 + {1 \over 4e}2^{-\left( {{N+2} \over 2}
\right)}{(N+2)^2 \over N+4}\left[1 - {4
\over {N+2}} + ...\right]\right\} $.
We also analyze the behavior for disorder with long-range
correlations, as is appropriate for interfaces in random field
systems, and study the crossover between the two regimes.  }
\vfill\eject\footline={\hss\tenrm\folio\hss}                              %
\vskip .3in
\newsec{Introduction and Summary}

Oriented elastic manifolds embedded in spaces which contain random
impurities that can pin the manifold occur in many physical systems.
Both the dimension of the manifold, $d$, and the dimension of the
space in which it is embedded, $D = d + N$, where $N$ is the number of
transverse dimensions, play important roles.\nref\rMP{M. Mezard and G.
Parisi, J.  de Phys.  {\bf I 1}, 809 (1991).}\nref\rHH{T.
Halpin-Healy, Phys.  Rev.  A {\bf 42}, 711 (1990).}\cite{\rMP,\rHH}
The case $d=1$ corresponds to a directed polymer in a random
potential\nref\rKZ{M.  Kardar and Y.-C. Zhang, Phys. Rev.  Lett. {\bf
58}, 2087 (1987).}\nref\rFH{D. S. Fisher and D. A. Huse, Phys. Rev. B
{\bf 43}, 10728 (1991).}\cite{\rKZ,\rFH}\ which describes the
interaction of a single flux line in a type-II superconductor with
impurities.\ref\rNS{D. R.  Nelson, Phys. Rev. Lett. {\bf 60}, 1973
(1988); D. R. Nelson and H. S.  Seung, Phys. Rev. B {\bf 39}, 9153
(1989).}\ Interfaces between two coexisting phases in $D$ dimensional
systems are $D-1$ dimensional oriented manifolds whose properties
control much of the behavior of such systems in the presence of
randomness.\nref\rIM{Y. Imry and S.-K.  Ma, Phys. Rev. Lett. {\bf 35},
1399 (1976); G. Grinstein and S.-K. Ma, Phys. Rev B {\bf 28}, 2588
(1983).}\nref\rZLR{G. Grinstein and J. Fernandez, Phys. Rev B{\bf 29},
6389 (1984); J. Villain, Phys. Rev. Lett. {\bf 52}, 1543
(1984).}\nref\rDSFFRG{D. S.  Fisher, Phys.  Rev.  Lett. {\bf 56}, 1964
(1986).}\cite{\rIM-\rDSFFRG}\ It has also been argued that flux
lattices in superconductors have an intermediate distance regime in
which they behave like a $d=N=3$ oriented manifold.\ref\rYB{J.P.
Bouchaud, M.  M\'ezard, and J. Yedidia, Phys.  Rev. Lett. {\bf 67},
3840 (1991).}

{}From a theoretical point of view, it has become clear in the last few
years that elastic manifolds in random media exhibit much of the
interesting and subtle phenomena that characterize spin-glasses and
other complicated disorder dominated phases.\nref\rFHSG{D. S.  Fisher
and D. A. Huse, Phys. Rev. B {\bf 38}, 386 (1988).}\cite{\rFH,\rFHSG}
In addition, the equilibrium statistical mechanics of directed
polymers in a random potential can be mapped to the dynamics of an
interface growing by random deposition -- leading to insights into
both problems.\ref\rKPZ{M. Kardar, G. Parisi, and Y.-C. Zhang, Phys.
Rev.  Lett. {\bf 56}, 889 (1986).}\ In this paper we will analyze the
equilibrium behavior of manifolds in random media for $d$ just below
the critical dimension of four, focusing on the limit of large $N$.

By definition, an oriented manifold has no overhangs; therefore, it
can be described entirely in terms of a set of $N$ transverse
coordinates $\{\phi^i \}$, which are functions of the $d$ internal
coordinates $\{ x^j \}$ parameterizing the manifold.\ref\fLatin{We use Latin
indices to denote both internal and transverse coordinates, reserving
Greek indices for replicas to be introduced later.}\ The Hamiltonian
describing such a manifold in the presence of a quenched random
potential $\vbare$ is
\eqn\eHambare{ H = \xint \left\{ {1 \over 2} \nabla\bphi \cdot
\nabla\bphi + \vbare. \right\},}
where $\bphi(\bx) \in \Re^N$ is a vector describing the transverse
coordinates of the manifold with internal coordinate $\bx \in \Re^d$.

The random potential and/or thermal fluctuations will generally cause
the manifold to roughen, resulting in divergent fluctuations in
$\bphi$.  It is conventional to parameterize these by a roughness
exponent $\zeta$:
\eqn\eZetadefinition{\left\langle\left\langle\left[ \bphi(\bx) -
\bphi(\bx')\right]^2 \right\rangle_{T}\right\rangle \sim |\bx -
\bx'|^{2\zeta},}
where the inner angular brackets with a $T$ subscript denote a thermal
average, while the outer brackets indicate a statistical average over
the random potential.

In the absence of the random potential, the interface will be flat for
$d>2$, corresponding to $\zeta=0$, but be thermally rough with
$\zeta=\zeta_{T}= (2-d)/2$ for $d \le 2$ (with logarithmic corrections
in two dimensions).  For $d<2$, weak randomness is irrelevant when
$N>N_T = 2d/(2-d)$, and an unpinned phase exists at high
temperature.\ref\rDF{C. A. Doty and D. S. Fisher, unpublished.} For a
strong random potential, low temperature or outside this regime of the
$N-d$ plane, the randomness always dominates and the system is
controlled by a non-trivial zero-temperature fixed point.

Much is known about the case of one internal dimension ($d = 1$),
owing both to the mapping to interface growth\cite{\rKPZ}\ and to the
simplicity of numerical simulations.\ref\rNumerics{For a review, see
J. Krug and H. Spohn in {\it Solids Far From Equilibrium: Growth,
Morphology and Defects} edited by C.  Godreche (Cambridge University
Press, New York 1990).} When both $d=1$ and $N=1$, the roughness
exponent is known to be exactly $\zeta=2/3$.\nref\rHHF{D. A. Huse, C.
Henley, and D. S. Fisher, Phys. Rev. Lett. {\bf 55}, 2924
(1985).}\cite{\rKZ,\rHHF} For $N \ge 2$, in addition to the low
temperature randomness--dominated phase, a high-temperature phase
emerges, in which the disorder is irrelevant and $\zeta = 1/2$. The
value of $\zeta$ in the low-temperature phase has been investigated
numerically for $N \ge 2$; it decreases with $N$ and appears to
approach $1/2$ as $N \goes \infty$.\cite{\rNumerics} The
possibility of a finite upper critical dimension, such that $\zeta =
1/2$ for $N>N_c$, has been suggested by several authors\nref\rDS{B.
Derrida and H. Spohn, J. Stat. Phys. {\bf 51}, 817
(1988).}\cite{\rHH,\rDS}, but no evidence of this has appeared in
rather extensive numerical simulations, and others have argued that no
such upper critical dimension exists, but rather that $\zeta$
decreases continuously to $1/2$ as $N \goes \infty$.\cite{\rFH}

For manifolds with $d > 4$, a perturbative analysis\cite{\rDSFFRG}\ shows
that the interface remains flat ($\zeta = 0$).  In this regime
perturbation theory, or equivalently a simple RG treatment (briefly
described in section II) is valid.  For $d<2$, as mentioned above,
both high temperature pinned and low-temperature randomness--dominated
phases exist for large enough $N$.  Between two and four dimensions,
only the disorder--dominated phase exists, characterized by a
non-trivial $\zeta$.  The first attempts to analyze this phase
involved perturbative methods about four dimensions -- e.g.  Parisi
and Sourlas
\ref\rPS{G. Parisi and N. Sourlas, Phys. Rev.  Lett. {\bf 43}, 744
(1979).}\ -- which yield incorrect results (in particular $\zeta =
(4-d)/2$) due to the existence of many extrema of the Hamiltonian
Eq.\eHambare.  In order to carry out a proper $\epsilon$ expansion, a
functional renormalization group (RG) is needed.  This was introduced
by Fisher\cite{\rDSFFRG}\ for the case $N=1$, yielding the results $\zeta =
\epsilon/3$ and $\zeta = 0.2083\epsilon$ for random-field and
random-bond disorder respectively to lowest order in $\epsilon=4-d$.
The first result\cite{\rZLR}\ is believed to be exact for $1<d<4$, while the
second is an
$O(\epsilon)$ calculation, which required a numerical solution of the
RG fixed point equation.

More recently, Mezard and Parisi\cite{\rMP}\ (MP) have performed an
approximate ``replica symmetry breaking'' calculation on the model for
general $N$, and argued that their results are exact in the limit as
$N \goes \infty$.  In MP's method, replica symmetry breaking is
introduced as a {\sl variational} ansatz.  An interesting question to
address is whether the replica symmetry breaking corresponds to some
physical aspect of the problem (at least at large $N$) or is merely a
feature of the restricted variational ansatz. This is a particularly
interesting issue in light of our $O(\epsilon)$ treatment, which {\sl
does not} involve replica symmetry breaking, but is a systematic
perturbative RG calculation.  It is hoped that a comparison of the two
methods may provide insight into whether ``replica symmetry breaking''
has any well-defined meaning.

In this paper, we generalize the $4-\epsilon$ RG calculation of
ref.\cite{\rDSFFRG} to arbitrary $N$.  In the limit of large $N$, the
fixed-point and stability equations become tractable, and we perform
an expansion around this limit, working always only to first order in
$\epsilon$.  For short-range correlated disorder, we find
\eqn\eRoughnessSR{\zeta_{SR} = {\epsilon \over {N+4}}\left\{1 + {1
\over
4e}2^{-\left( {{N+2} \over 2} \right)}{(N+2)^2 \over N+4}\left[1 - {4
\over {N+2}} + ...\right]\right\},}
while for a random potential with long-range correlations transverse
to the manifold,
\eqn\eLongrangecorrelations{\langle V(\bphi,\bx)V(\bphi',\bx')
\rangle_C \sim |\bphi - \bphi'|^{-\gamma}
\delta^{d}(\bx - \bx'),}
there is a continuously variable exponent which is independent of $N$
\eqn\eRoughnessLR{\zeta_{LR} = \epsilon/(4+\gamma),}
in agreement with previous results for the particularly interesting
case of interfaces in random-field systems which corresponds to $N=1$
and $\gamma = -1$.  The long-range fixed point becomes unstable when
$\zeta_{LR} < \zeta_{SR}$.  As in the $N=1$ case, the long-range
result is believed to be exact, while the short-range result is true
only to $O(\epsilon)$.

Before proceeding with the RG calculation,we briefly outline the
remainder of the paper.  In the next section, the model is described,
and an RG procedure is developed to analyze the zero temperature fixed
point in $4-\epsilon$ dimensions.  It is shown that a perturbative
expansion, which incorrectly deals with the physics of the many
metastable states, breaks down.  Section III analyzes the correct
solution for the fixed point and roughness exponent in the large $N$
limit, in which analytic results can be obtained.  The stability of
this fixed point is analyzed in section IV.  The behavior for
long-range correlated disorder (e.g. random fields) is studied in
section V, as well as the stability of the associated fixed points.
In section VI we summarize our conclusions and open questions and
suggest several possible directions for future work.  Appendices A and
B contain various technical details, while Appendix C rederives the RG
relations by an iterative minimization of the Hamiltonian and
discusses the appearance of many minima.  Finally, Appendix D analyzes
possible multicritical fixed points.

\newsec{Model and Zero-Temperature Renormalization Group}

The partition function in terms of the Hamiltonian Eq.\eHambare\ is
\eqn\eZofv{ Z\{V\} = \int\! [d\bphi]\; \exp ( - H/T ).}
We take the random potential $V$ to have a Gaussian distribution, with
the two point correlation function
\eqn\eVtwopoint{ \left\langle V(\bphi,{\bf x})V(\bphi',{\bf x}')
\right\rangle = R(\bphi-\bphi')\delta^{(d)}({\bf x - x'}),}
with a short-distance scale implicitly included in the
$\delta$-function.  To order $\epsilon = 4 - d$ higher cumulants will
be shown to be irrelevant.

To organize a renormalization group (RG) treatment of the problem, we
employ the replica method of averaging over the disorder.  Note,
however, that we do this only to organize the perturbation expansion.
The partition function is now a random variable, and rather than
follow the flow of its distribution function directly under the RG,
one can follow the full set of moments.  These are readily averaged
over, yielding terms of the form
\eqn\eZreplica{\overline{Z^p} = \int\! [d\bphi] \; \exp ( - \tilde{H}_p ), }
with
\eqn\eHamreplica{\tilde{H}_p \!\! = \!\! \xint \!\! \left\{ {1 \over {2T}}
\sum_\alpha
\nabla\bphi^\alpha\! \cdot
\nabla\bphi^\alpha\! - {1 \over {2T^2}} \sum_{\alpha\beta}
R(\!\bphi^\alpha\!\! -\! \bphi^\beta\!) - {1 \over T^3}
\sum_{\alpha\beta\gamma} S(\!\bphi^\alpha\!\! -
\!\bphi^\beta\!,\!\bphi^\alpha\!\! - \!\bphi^\gamma\!) - \cdots
\right\},}
where the terms with three or more replicas result from non-Gaussian
correlations in the disorder; these are generated at higher order in
$\epsilon$. We have not included terms with additional gradients,
which will be irrelevant. The symmetry under simultaneous shifts of
all the replica coordinates corresponds to the statistical symmetry of
the disorder under spatial translation normal to the manifold.  We
have used $p$ instead of the more conventional $n$ for the number of
replica indices to avoid confusion with the transverse dimension $N$
of the manifold.

The momentum shell RG approach we employ consists of integrating out
high-momentum modes in a shell with $\Lambda/b < |{\bf p}| <
\Lambda$.  We take the limit in which the width of this shell is
infinitesimal, i.e. $b = e^{dl}$, which simplifies the formulae
somewhat.  To keep the cut-off fixed, momenta, coordinates, and fields
are rescaled according to
\eqn\eRescale{\eqalign{
{\bf p} = & \; {\bf p}'\!/b, \cr
\bx = & \; b\bx', \cr
\bphi(\bx) = & \; b^\zeta \bphi'(\bx'). \cr}}

A simple first attempt at an RG analysis consists of expanding the
function $R(\bphi)$ in a power series, and analyzing the results term
by term.  Since $R$ is an even function, only even powers of $\bphi$
appear in such an expansion, i.e.
\eqn\eRexpansion{\tilde{H}_{int} = -{1 \over
{2T^2}}\sum_{\alpha,\beta}\sum_m {R_m
\over m!}
(\bphi^\alpha - \bphi^\beta)^{2m}.} Above four dimensions, the quartic
and higher vertices are irrelevant and the theory flows to a simple
Gaussian fixed point.  Below four dimensions the quartic interaction
becomes relevant, and one might hope to make a simple epsilon
expansion by going to second order.  A simple calculation shows that
no perturbative fixed point exists, so that a strong-coupling analysis
is necessary.

Physically, the behavior below four dimensions is dominated by the
randomness, and should be described by a zero temperature fixed point.
By allowing temperature to renormalize, it is possible to organize an
analysis of this fixed point.  The RG flows which arise from the scale
changes are then
\eqn\ePCflow{\eqalign{
\left.{{dT} \over {dl}}\right|_{SC} = & \; \; (2 - d - 2\zeta)T , \cr
\left.{{\del R(\bphi)} \over {\del l}}\right|_{SC} = & \; \; (4 - d -
4\zeta)R(\bphi) +
\zeta \phi_i \del_i R(\bphi) ,\cr
\left.{{\del S(\bphi_1,\bphi_2)} \over {\del l}}\right|_{SC} = & \; \; (6 -
2d - 2\zeta)S(\bphi_1,\bphi_2) + \zeta(\phi_{1i}\del_{1i} +
\phi_{2i}\del_{2i})S(\bphi_1,\bphi_2), \cr
\cdots & \cr}}
The $\zeta\phi^i\del_i R(\bphi)$ term in the second equation comes
from the infinitesimal field rescaling of Eq.\eRescale.

The flow equation (Eq.\ePCflow) for the temperature is actually exact,
due to the ``Galilean'' invariance of the Hamiltonian Eq.\eHambare.
This can be seen by considering the change in the free energy
corresponding to a uniform tilt of the manifold (or equivalently a
change in boundary conditions).  If the fields are shifted by a linear
function of the coordinates $\bphi
\goes \bphi + {\bf v}^i x^i$, the probability distribution of the new
free energy is identical to that of the old one plus an additive
constant
\eqn\eFreeeenergychange{\Delta F \equiv \Delta \left( -T\ln Z \right)
= {1 \over 2} |{\bf v}|^2 L^d.} This is an exact statement about the
model, and must therefore be true at all stages of the RG; it requires
that $T$ (i.e. the coefficient of the stiffness term) only be
renormalized by the scale changes.\ref\rSchulz{U. Schulz, J. Villain,
E. Br\'ezin, and H. Orland, J. Stat. Phys. {\bf 51}, 1 (1988).} We
thus have
\eqn\eTscaling{{dT \over {dl}} = -\theta T,}
with $\theta = 2\zeta +d -2$ determining the scaling of energies at
the zero temperature fixed point.

One can again attempt to proceed by expanding the function $R(\bphi)$
in a power series.  The condition fixing the quadratic term $R_2$ in
Eq.\eRexpansion\ is then $\zeta = (4-d)/2 = \epsilon/2$.  However,
since $\zeta$ is of order $\epsilon$, all the higher terms in the
expansion of $R(\bphi)$ also become relevant below four dimensions.
It is therefore necessary to keep track of the entire series of $\{
R_m \}$, or the entire function $R(\bphi)$.

The one-loop RG equations are best derived in their functional form,
through the use of Fourier transforms.  As an example, consider the
{\sl first order} feedback of the $R(\bphi)$ term.  As noted above, it
will not renormalize itself, but it will contribute to the free
energy.  The first step in the calculation is to Fourier transform the
interaction term,
\eqn\eFTterm{-\tilde{H}_{int}  = {1 \over {2T^2}}\sum_{\alpha,\beta}
\int_{\bx,\bk} e^{i\bk\cdot (\bphi^\alpha (\bx) -
\bphi^\beta (\bx) )} \tilde{R}(\bk),}
where we use Greek and Latin letters for Fourier transforms
perpendicular ($\bphi$ direction) and parallel ($\bx$ direction) to
the manifold, respectively, and $\int_{\bx,\bk} \equiv \int d^d
\!{\bx} d^d  \!\bk/(2\pi)^d$.

To perform the elimination part of the RG transformation, the fields
are split into slowly and rapidly moving parts,
\eqn\ePhisplit{\bphi(\bx) = \bphi_<(\bx) + \bphi_>(\bx),}
and a trace is performed over the ``fast'' fields $\bphi_>(\bx)$.  In
Eq.\eFTterm, the slow and fast terms separate into two exponential
factors, and it is a simple matter to average over the fast modes.
Dropping the label ($<$) on the slow fields, the traced term becomes
\eqn\eFORG{-\left\langle \tilde{H}_{int} \right\rangle_> = {1 \over
{2T^2}}\sum_{\alpha,\beta} \int_{\bx,\bk}e^{i\bk\cdot (\bphi^\alpha
(\bx) -
\bphi^\beta (\bx) )} e^{-\kappa^2 T G_>(0)(1 - \delta^{\alpha\beta})}
\tilde{R}(\bk),}
where the free two-point function is
\eqn\eGdef{G_>(\bx) = \int_{\bf q}^{>} e^{i{\bf q}\cdot\bx}/q^2 , }
where the $>$ on the integral denotes integration over momenta (${d^d
{\bf q}/(2\pi)^d}$) in the shell only.  When evaluated at $\bx = 0$,
this function has the well-defined limit
\eqn\eGzero{G_>(0) = (2\pi)^{-d}S_d \Lambda^{d-2} dl \equiv A_d dl,}
where $S_d$ is the surface area of a unit sphere in $d$ dimensions.
The second exponential in Eq.\eFORG\ can thus be expanded to yield
\eqn\eFORGtwo{-\left\langle \tilde{H}_{int} \right\rangle_> = {{A_d dl}
\over
{2T}} \int_\bx \left\{\sum_{\alpha,\beta} \del_i\del_i R(\bphi^\alpha
-
\bphi^\beta) - p \del_i\del_i R(0)\right\},}
where the $\del_i$ act on the internal coordinates of the functions
$R(\bphi)$ (not on the spatial coordinates $\bx$) and repeated Latin
indices are summed from $1$ to $N$.  The first term is of the
appropriate form to feed back into $R(\bphi)$, but it is reduced by a
factor of $T$ from the term in Eq.\eHamreplica.  Physically, this
renormalization is due to the averaging of the potential by thermal
fluctuations, and is thus negligible at the zero temperature fixed
point of interest.  The second term in Eq.\eFORGtwo\ contributes to
the renormalization of the free energy density
\eqn\eRGfree{{df \over dl} = (d-\theta)f - {A_d \over 2} \del_i\del_i R(0)
+ \cdots.}

We now proceed with the analysis to second order in $R(\bphi)$; one
must multiply two terms and take the connected expectation value over
$\phi_>$.  Performing the average, one finds the expression
\eqn\eRGsecondorder{\eqalign{{1 \over
{8T^4}}\sum_{\matrix{\scriptstyle
\alpha_1,\beta_1 \cr \scriptstyle \alpha_2,\beta_2 \cr}}
\int_{\matrix{\cr \scriptstyle \bx_1\bk_1 \cr
\scriptstyle \bx_2, \bk_2}}\!\!\! &
 \exp\left\{ {i\bk_1\cdot (\phi^\alpha_1 (\bx_1) -
\phi^\beta_1 (\bx_1) )+ i\bk_2\cdot (\phi^\alpha_2 (\bx_2) -
\phi^\beta_1 (\bx_2) )} \right\}\cr \times & \exp\left\{
 -\kappa_1^2 T G_>(0)(1 -\delta^{\alpha_1\beta_1}) -\kappa_2^2 T
G_>(0)(1 - \delta^{\alpha_2\beta_2})\right\}
\tilde{R}(\bk_1)\tilde{R}(\bk_2) \cr
\times & \exp \left\{
-\bk _1 \cdot \bk_2 T G_>(\bx_1 - \bx_2)\left[
\delta^{\alpha_1\alpha_2} +
\delta^{\beta_1\beta_2} - \delta^{\alpha_1\beta_2} -
\delta^{\beta_1\alpha_2}
\right] \right\} \cr & - {\rm Disconnected} \hskip 0.1 in {\rm Parts}.
\cr}}
The terms resulting from expanding the $G_>(0)$ parts are canceled by
disconnected pieces, leaving only the expansion of the final term.
Expansion of the final exponential gives terms proportional to $1/T^3$
and lower powers of $T$.  The $1/T^3$ term is a three-replica
contribution,
\eqn\eThreereplica{{1 \over {2T^3}}\sum_{\alpha,\beta,\gamma}
\int_{\bx,\bx'} \del_i R(\bphi^\alpha\!(\bx) - \bphi^\gamma\!(\bx))
\del_i R(\bphi^\alpha\!(\bx')- \bphi^\beta\!(\bx')) G_>(\bx - \bx'),}
since a single Kronecker delta function leaves three free replica
indices.  Eq.\eThreereplica, however, contributes only at large
momenta, since $G_>(\bx)$ contains an integral only over momenta in
the shell.  Because the generated 3-replica term exists only at large
momentum, it cannot feed back to generate a zero-momentum term until
second order.  The resulting contribution to the renormalization of
$R$ will then turn out to be higher order in $\epsilon$, since the
3-replica piece will be $O(\epsilon^2)$, anticipating that $R$ will be
$O(\epsilon)$ at the fixed point.  The situation is analogous to the
neglect (to lowest order) of the $\phi^6$ term in momentum-shell RG
for the conventional $\lambda\phi^4$ theory, in which a $\phi^6$
interaction with large momentum is generated from two $\phi^4$ terms,
but does not feed back in a dangerous manner.  Fig.[1] shows
diagrammatically how the three-replica term is generated and feeds
back into the two-replica piece.

The $1/T^2$ parts generate both two- and three-replica terms. The
three-replica term is down by a factor of $T$, and can be neglected at
the zero temperature fixed point.  Keeping track of the factors, one
finds a contribution
\eqn\eRGcontrib{\eqalign{ {{\del R(\bphi)} \over {\del
l}}\bigg|_{O(R^2)} = & {1 \over {2T^2}}\sum_{\alpha,\beta}
\int_{\bx,\bx'} G_>(\bx - \bx')^2\bigg[
\del_i\del_j R(\bphi^\alpha(\!\bx\!)\! -\!
\bphi^\beta(\!\bx\!))\del_i\del_j R(\bphi^\alpha(\!\bx'\!)\!  -\!
\bphi^\beta(\!\bx'\!)) \cr - & 2\del_i\del_j
R(\bphi^\alpha(\!\bx\!)\!-\!\bphi^\beta(\!\bx\!)) \del_i\del_j R(0)
\bigg].\cr}}
Since we are interested in the renormalization of the long-wavelength
portion, the kernel $K(\bx) \equiv G_>(\bx)^2$ must be evaluated at
zero momentum, i.e.
\eqn\eKernel{\tilde{K}({\bf p = 0}) = \int_{\bx} G_>(\bx)^2 =
\int_{\bx,{\bf p,p'}}^> {{\exp\left[i{\bf (p + p')}\cdot\bx\right]}
\over {\bf p^2p'^2}} = \int_{\bf p}^{>} {1 \over {\bf p^4}} = S_d (2\pi)^{-
d}
\Lambda^{d-4} dl.}
By rescaling $R(\bphi)$ by a constant multiple, this factor can be
removed.  The full RG equation for $R$ then becomes
\eqn\eRGR{\eqalign{ {{\del R(\bphi)} \over {\del l}} = & (4-d-
4\zeta)R(\bphi) + \zeta
\phi^i\del_i R(\bphi) + \left[ {1 \over 2}\del_i\del_j R(\bphi)
\del_i\del_j R(\bphi) - \del_i\del_j R(\bphi) \del_i\del_j R(0)
\right] \cr
 & + O(R^3).\cr}} This equation can also be formally derived by
expanding $R(\bphi)$ in $\bphi_>$ and keeping terms up to
$O(\bphi_>^4)$, although such a treatment does not properly treat the
case of non-analytic $R(\bphi)$.  The flow equation Eq.\eRGR\ is
exactly equivalent to the infinite series of RG equations obtained
from all one loop diagrams in an ordinary diagrammatic approach.  It
was used previously in Ref.\cite{\rHH} and Ref.\ref\rNL{T. Natterman
and H. Leschhorn, Europhys. Lett. {\bf 14}, 603 (1991).}.  In
Appendix C, we derive the RG equation schematically by directly
minimizing $H$ over the fast degrees of freedom $\bphi_>$ without the
use of replicas or field theoretic techniques.

At this point, one may try to directly analyze the flows and fixed
points of Eq.\eRGR: if a fixed point $R \sim \epsilon$ is found, then
the other terms in $\tilde{H}_p$ will not play a role to
$O(\epsilon)$.  We will take this approach in the next section, but
first it is instructive to investigate the manner in which the
standard polynomial RG breaks down.  A conventional RG approach is
equivalent to expanding $R(\bphi)$ in a power series
\eqn\eRpowerseries{R(\bphi) = \sum_m {R_{2m} \over (2m)!} \phi^{2m},}
and following the flow equations for the coefficients.  The first two
equations are
\eqn\ePSflow{\eqalign{
{{\del R_2} \over {\del l}} = & (\epsilon - 2\zeta)R_2 , \cr {{\del
R_4} \over {\del l}} = & \epsilon R_4 + {{N+8} \over 3}R_4^2.
\cr}}
While the quadratic term could be fixed by requiring $\zeta =
\epsilon/2$ (the naive perturbative result), $R_4$ then flows off and is not
stabilized at second order.  Examination of the flow equation for
$R_4$ shows that it becomes infinite after a finite amount of
renormalization.  This is an artifact of the truncation to second
order; nevertheless, $R_4$ will rapidly become $O(1)$.

We thus see that there is no fixed point of the perturbative RG for
which $R(\bphi)$ is analytic.  It is the assumption of analyticity
which leads to this conclusion, and we shall see that non-analytic
fixed points of Eq.\eRGR\ do exist.  The correct behavior for small
$\bphi$ can be found by directly examining Eq.\eRGR\ with $\del R/\del
l = 0$.  This will be done in the next section.

Before proceeding, it is useful to recall how a normal perturbative
analysis leads to drastically wrong results.  Various methods have
been used to show that, in a field theoretic expansion, formally $R_2$
has no non-trivial renormalizations.\cite{\rPS}\  An apparent fixed point can
then
be found by setting $\zeta=\epsilon/2$ to all orders in $\epsilon$.
This results in the so-called ``dimensional reduction'' result $\theta
= 2$.  In the supersymmetric formulation of Parisi and Sourlas which
averages over {\sl all} of the extrema of $H$ with positive and
negative weights, the only operator which appears corresponds to
$R_2$.  Thus this analysis completely misses the flow of the other
operators, such as $R_4$, out of the regime in which a perturbative
analysis might have been valid.  As we discuss in the conclusion, it
is likely that there is an exact upper bound for $\zeta$ which is
violated by the naive perturbative result.

\newsec{Fixed Point Analysis in the Large $N$ Limit}

In this section, we analyze the behavior of the fixed points in
$4-\epsilon$ dimensions in the limit of large $N$.  We look for
solutions where the $R$ is $O(1/N)$ by rescaling $R
\rightarrow R/N$.  The flow equation (Eq.\eRGR) then takes the form
\eqn\eFlowR{ {{\del R(\bphi)} \over {\del l}} = (4-d-4\zeta)R(\bphi) +
\zeta
\phi^i\del_i R(\bphi) + {1 \over N}\left[ {1 \over 2}\del_i\del_j R(\bphi)
\del_i\del_j R(\bphi) - \del_i\del_j R(\bphi) \del_i\del_j R(0) \right].}

For rotationally invariant solutions, the ansatz $R(\bphi) =
Q(\phi^2/2)$ yields
\eqn\eFlowQ{{{\del Q} \over {\del l}} = (4-d-4\zeta)Q + 2\zeta y Q' +
{1 \over 2}(Q')^2 - Q'Q'_0+ {1 \over N}\left[2yQ'Q'' + 2y^2(Q'')^2 -
2yQ''Q'_0 \right],} where primes denote differentiation with respect
to $y$, defined as
\eqn\eYdef{y \equiv \phi^2/2,} and
\eqn\eQzerodef{Q'_0 \equiv Q'(y=0).}
Differentiating once and pulling out the $O(\epsilon)$ factor by
defining
\eqn\ePullitout{\eqalign{
Q' = & -(4 -d -2\zeta){N \over {N+2}} u, \cr {\rm and}\;\; t = & (4 -
d - 2\zeta)l, \cr}} gives the final form of the fixed point equation:
\eqn\eFPfull{ {{\del u} \over {\del t}} = u + \beta y u' + u'(u(0) - u) -
\mu\left[ 3y (u')^2 +
2y^2 u'u'' + yu''(u - u(0))\right] = 0,} with
\eqn\eBetamudef{\eqalign{
\beta \equiv  \; \; & 2\zeta/(4-d-2\zeta), \cr
\mu \equiv \; \; & 2/(N+2). \cr}}
Any physical solution of this equation will have a finite value of $u$
at the origin, so that by a choice of scale we can set
\eqn\eUzerodef{u(0) =1.}
The behavior at large $y$ will fix the value of $\beta$ in a way which
is somewhat analogous to more conventional eigenvalue problems.  If we
start with potential correlations which are short range, it is natural
to expect the fixed points also to correspond to short range
correlations.  We thus look for fixed points with $R(\bphi)$ decaying
rapidly for large $\bphi$; i.e. $u(y)$ decaying rapidly for large $y$.

Naively, in the large $N$ limit the terms proportional to $\mu$ in
Eq.\eFPfull\ can be simply dropped, and the resulting problem has a
much simpler form:
\eqn\eNaive{ u_\infty + \beta y u^\prime_\infty + u^\prime_\infty(1 -
u_\infty) = 0.} We will see that this approximation is not valid
globally, but this equation determines the primary solution, called
the outer solution in the language of boundary-layer
theory.\ref\rBO{C. Bender and S.  Orszag,{\it Advanced Mathematical
methods for Scientists and Engineers,} (McGraw-Hill, NY, 1978).} This
is in a sense the solution for $N=\infty$.  The boundary-layer, or
region in which this solution is not valid, occurs for large $y$.  The
solution in this tail region is conventionally denoted the inner
solution (as it is valid inside the boundary layer).  The primary
limit (Eq.\eNaive) has only power-law solutions at infinity unless
$\beta = 0$. We thus tentatively choose this value (anticipating later
corrections for large but finite $N$). In this case the solution to
Eq.\eNaive\ is given implicitly by
\eqn\eImplicit{u_\infty - \log u_\infty = y + 1,}
using the boundary condition $u(0) = 1$.  For small $y$, this equation
yields two possible behaviors for $u(y)$,
\eqn\eSmallybehavior{u(y) \approx 1 \pm \sqrt{2y}.}   We
 choose the minus solution to obtain a solution decaying as $\exp(-y)$
for large $y$.  The value $\beta = 0$ corresponds to $\zeta = 0$, so
that the interface remains flat to $O(\epsilon)$ at $N=\infty$.  A
more careful analysis yields the detailed form of the corrections.

To expand towards finite $N$, the natural first step is to examine
perturbatively the effect of the terms dropped in Eq.\eNaive.  Since
the zeroth order solution is exponential for large $y$, it is
immediately clear that perturbation theory breaks down in this regime,
due to the presence of the term
\eqn\eBigterm{\eqalign{
 \mu y u'' \gg u \hskip 0.1 in {\rm for} & \; \; y \gg 1/\mu,\cr {\rm
with} & \;\; u \sim \exp(-y).}} The solution Eq.\eImplicit\ is
therefore valid (even approximately) only for $y \ll 1/\mu$, and we
thus have a boundary layer for $y \gg 1/\mu$.

Fortunately, we can proceed with the analysis by noting that for $y
\gg 1$, the non-linear terms in Eq.\eFPfull\ become negligible, and
the equation can be reduced to the form
\eqn\eLinearized{ u + \beta y u' + u' + \mu y u'' = 0,}
where we have allowed for $\beta \neq 0$ (but $O(\mu)$), which will
turn out to be the case for for $N$ large.  Eq.\eLinearized\ is valid
within the boundary layer.  It is important to note that both this
equation for the tail and the primary equation (Eq.\eNaive) are valid
in the (asymptotically infinite) region $1 \ll y
\ll 1/\mu$, which makes it possible to match the solutions of the two
equations in this domain (see Fig.[2]).

The linear tail equation (Eq.\eLinearized) is second order, and
therefore has two independent solutions.  For $\mu y, \beta y \ll 1$,
the solutions are
\eqn\eRightmatch{\eqalign{
u_a(y) \sim & \; \; Ce^{-y}, \cr u_b(y) \sim & \; \; Cy^{-1/\mu}.
\cr}} From the behavior of the primary solution, Eq.\eImplicit, in the
matching region $y \gg 1$, we see that we must choose the solution
$u_a(y)$ (with possibly a small admixture of $u_b$ vanishing in the $N
\goes \infty$ limit).  For a given value of $\beta$, both the primary and tail
solutions are thus completely determined, giving a uniformly
valid solution to the full equation (Eq.\eFPfull) for large $N$.

For short-range correlated disorder, the bare unrenormalized function
$u(y)$ decays exponentially (or more rapidly) for large $y$.  It is
straightforward to see that this exponential decay is preserved by the
flows, from Eq.\eFPfull, or directly from Eq.\eFlowR.  Higher order
terms from higher-loops can generate at most power-law corrections to
the initial exponential behavior (from terms with all but one of the
$R^{'}$s evaluated at $\bphi = 0$).  The non-renormalization of
the exponential behavior can be seen in a schematic way directly from
the Hamiltonian.  When a particular fast mode is integrated out, as in
the previous section, the integral performed is of the form
\eqn\eSchematicint{ \int [d\bphi_>] \exp \left[ -\bphi^2_>/2+
R(\bphi_< \! + \!\bphi_>) \right].} For large $\bphi_<$, the
contributions from small and large $\bphi_>$ can be easily estimated.
For small $\bphi_>$, the argument of $R$ is large, so that it can be
approximated by an exponential decay $R \sim
\exp( - \bphi_<^2 /2)$, and the integral yields an exponentially
decaying function of $\bphi_<$.  For large $\bphi_>$, the integral is
dominated by $\bphi_> \approx - \bphi_<$, so that now the quadratic
term dominates and again yields an exponential function of $\bphi_<$,
of the exact form that corresponds to $u(y) \sim \exp(-y)$.

To be a valid fixed-point function for short-range correlated
disorder, therefore, $u(y)$ must have an exponential tail at large
$y$.  By analyzing the tail equation (Eq.\eLinearized) for $y \gg
1/\beta, 1/\mu$, we again find two possible behaviors:
\eqn\eLargey{\eqalign{
u_c(y) \sim & \; \; Cy^{1/\beta-1/\mu}e^{-(\beta/\mu)y}, \cr u_d(y)
\sim & \; \; Cy^{-1/\beta}. \cr}} For most values of $\beta$, the
required solution $u_a(y)$ in the intermediate region $1\ll y \ll
1/\mu,1/\beta$ will be a linear combination of $u_c(y)$ and $u_d(y)$,
and thus will have a power-law tail at large $y$.  For some special
values of $\beta$, however, $u_a(y)$ will correspond exactly to
$u_c(y)$, so that the power-law tail of $u_d(y)$ does not contribute.
This will be the eigenvalue-like condition that determines $\beta$,
and hence the short-range roughness exponent $\zeta_{SR}$.

It is, in fact, simple to guess one such value of $\beta$.  From the
fact that $\beta$ vanishes for $N=\infty$ ($\mu=0$), one expects that
$\beta = O(\mu)$.  It is easy to check that for $\beta = \mu$, an
exact solution of Eq.\eLinearized\ satisfying the matching conditions
is $u = e^{-1-y}$. This value of $\beta$ corresponds to a roughening
exponent of
\eqn\eFirstapproxzeta{\zeta \approx \epsilon/(4+N),}
which we anticipate will be valid for large
$N$.\ref\fMulticritical{Actually, Eq.\eLinearized\ has a discrete
family of eigenfunctions satisfying the appropriate boundary
conditions for a short-range fixed point function and matching to the
primary solution.  In the linearized approximation to which we have
worked so far, these can all be found explicitly.  This calculation is
performed in appendix D, where it is shown that these may describe
multi-critical points for special correlations of the disorder.} Note
that this agrees with a recent replica symmetry breaking calculation,
which was claimed to be valid in the large $N$ limit.\cite{\rMP}

To compute the next order corrections to the result
(Eq.\eFirstapproxzeta), we need to consider in detail the effects of
the neglected terms.  For the primary solution ($y \ll 1/\mu$), these
can be computed perturbatively in $\mu$.  However, they will not
affect $\beta$ unless the non-linearities for $y \gg 1$ are also taken
into account, since the solution found above can be scaled to match
the corrections to the primary solution.  We therefore first consider
the effects of non-linearities for $y \gg 1$.

To look for solutions close to the original one ($u_\infty$ in
Eq.\eImplicit), we make the change of variables
\eqn\eChangeofvars{\eqalign{
u = & \; \; \exp\left( -y - \int^y_0 \! dy' \sigma(y')\right), \cr
\beta = & \; \; \mu(1+b), \cr
\mu y = & \; \; \eta, \cr}}
where the final change of variables was made to concentrate on the
change of character of the solution for $y \sim 1/\mu$.  With some
algebra, Eq.\eFPfull\ is transformed to
\eqn\eTransformed{ \left(-b + {{\eta-1}\over \eta}\right)\sigma - b
+ \sigma^2 - \mu{d\sigma \over d\eta}+ g(\eta) = 0,} where
\eqn\eGdef{g \equiv {{\exp\left(-y-\int^y_0 \sigma\right)}  \over \eta}
\left\{(1+\sigma) - 4\eta(1+\sigma)^2 + 2{\eta^2 \over
\mu}(1+\sigma)^3 + \left[2\eta^2(1+\sigma) +\mu\eta\right]{d\sigma
\over {d\eta}}\right\}.}

We anticipate that $\sigma(\eta)$ will be exponentially small (in
$1/\mu$) for $\eta \sim 1$.  Then we can ignore terms of order
$\sigma^2$ and also terms of order $\sigma(\eta)\exp(-\eta/\mu)$.  We
see, however, that $g(\eta)$ has a part which depends on the behavior
of $\sigma(\eta)$ for $\eta$ of order $1/\mu$ (i.e. $y$ of order one)
via $u$, Eq.\eChangeofvars.  We can obtain this by perturbing in
$\beta$ and $\mu$ about the solution $u_\infty$ which is good in this
region.  Anticipating that $b \ll 1$, a perturbative calculation
performed in appendix A yields:
\eqn\eUapprox{ u \approx u_\infty +\mu u_1 \approx  e^{-1-y}\left(1 -
\mu +  O(\mu^2)\right) \quad {\rm for} \;\; y \gg 1.}
This expression is valid for $\eta \ll 1$ but we will see that the
corrections embodied in Eq.\eChangeofvars\ will be small for the
desired full solution.  From Eq.\eUapprox, the $g$ term becomes
\eqn\eGsubst{g = {e^{-1 - \eta/\mu} \over \eta}\left[
{2\eta^2 \over \mu} + 1 - 4\eta - 2\eta^2 + O(\mu)\right].}

Neglecting the $\sigma^2$ (and $\sigma\exp(-\eta/\mu)$) terms in
Eq.\eTransformed, we obtain a linear homogeneous equation for
$\sigma$.  This can be solved straightforwardly by introducing the
integrating factor
\eqn\eIntfact{F(\eta) \equiv \exp\left[- {1 \over \mu}\int^\eta \!
d\tilde{\eta}
\left({{\tilde{\eta} - 1} \over \tilde{\eta}} - b\right)\right] = \exp\left\{
{1 \over
\mu}\left[ (b-1)\eta + \log\eta\right]\right\},}
whence
\eqn\eSigmasolution{
\sigma(\eta) \approx {1 \over {\mu F(\eta)}} \left\{
C + \int_0^\eta d\tilde{\eta} \; F(\tilde{\eta}) \left[g(\tilde{\eta})
-b\right]\right\}.} For $\eta \goes 0$, this must match onto the
solution Eq.\eUapprox\ implying that, since $F(\eta)$ vanishes for
small $\eta$, the integration constant $C$ must be zero.  At the other
end, since $F(\eta)$ also vanishes for large $\eta$, $\sigma$ will
diverge unless the integral in Eq.\eSigmasolution\ is zero.  We thus
obtain the integral condition
\eqn\eIntcond{\int_{0}^{\infty} \!d\eta \; \left[b - g(\eta)\right]
e^{-{1 \over \mu}\left[\eta - \log\eta\right]} = 0.}

With the condition of Eq.\eIntcond, Eq.\eSigmasolution\ yields
\eqn\eSigmalargeeta{\sigma(\eta) \approx {b \over {\mu F(\eta)}}
\int_\eta^\infty d\tilde{\eta} \; F(\tilde{\eta}) \approx {b \over {1-b}}
+ {b \over {\eta(1-b)^2}} + O(\eta^{-2}),} for $\eta \gg 1$ and hence
\eqn\eUlargeeta{u(y) \sim y^{-{b \over {\mu(1-b)^2}}}\exp\left[ -
y\left(1 + {b \over {1-b}} \right)\right],} for $y \gg 1/\mu$.  Up to
corrections of order $b^2$ in $\sigma$ (which arise from the neglected
terms in Eq.\eTransformed) this agrees with the behavior of $u_c(y)$
from Eq.\eLargey.  We have thus found the desired exponentially
decaying full solution valid for the full range of
$y$.\ref\fExtremeBL{Note that for extremely large $y \gg 1/b^2$, this
solution is not uniformly good; however, to obtain this regime we can
easily match onto the known simple behavior of $u_c(y)$ for $1/\mu \ll
y \ll 1/b^2$}

As $\mu \goes 0$, both terms in Eq.\eIntcond\ can be evaluated by
steepest descents.  Expansion of each term around the saddle points
$\eta^* \in \{1, 1\!/\!2\}$ yields
\eqn\eBsolution{b = {1 \over 2e}2^{-{1 \over \mu}}{1 \over \mu}[1 -
2\mu + O(\mu^2)],} which is exponentially small for small $\mu$,
justifying our approximations.  We are now in a position to obtain the
roughness exponent from Eq.\eBsolution.  For large $N$, we have
\eqn\eRoughfinalSR{\eqalign{
\zeta = & {\epsilon \over {N+4}}\left( 1 + {b \over{1 + \mu}} \right) \cr
 = & {\epsilon \over {N+4}}\left\{1 + {1 \over 4e}2^{-\left( {{N+2}
\over 2} \right)}{(N+2)^2 \over N+4}\left[1 - {4
\over {N+2}} + ...\right]\right\}.\cr}}
Note that an {\sl approximate} analysis of the RG flow (Eq.\eFlowR) by
Natterman and Leschhorn\cite{\rNL}\ also gave the same prefactor, but
a different exponentially small correction.
If we naively truncate the series after the ``$1$'' in the last
bracket, we obtain for $N=1$
\eqn\eBold{\zeta(N=1) = {\epsilon \over 5}\left( 1 + 0.0585 \right) \approx
0.2117\epsilon,}
It is clear from the increasing nature of the next term that the
series is asymptotic; nevertheless, the magnitude of the correction to
$\epsilon/5$ compares fairly well with the direct numerical solution
of Eq.\eFlowR\ for $N=1$, which yielded $\zeta = 0.2083\epsilon$.

In appendix D, we show that in addition to the fixed point found in
this section, there is a discrete infinite series of fixed points with
$u(y)$ changing sign but still decaying rapidly.  At this point,
whether these are physically meaningful is unclear.

\newsec{Stability Analysis}

In this section, we analyze the stability of the short-range fixed
point found in the previous section to perturbations with both short
and long-range correlations.  Since this fixed point represents a {\sl
phase} of the system, rather than a critical point, there should be no
short-range perturbations which are relevant.  From the scale
invariance of the fixed-point equation, it is clear, however, that
there is at least a marginal operator connecting the line of
short-range fixed points with different $u(0)$.  This does not alter
the physics, since it just corresponds to redefinitions of dimensional
quantities and is thus an uninteresting {\sl redundant} operator.
Intuitively, one expects perturbations with sufficiently long-range
correlations to be relevant, causing the system to flow to an
appropriate long-range fixed point; we will see that this is indeed
the case.

We look for the eigenoperators in the usual way, by considering a
function $u(y)$ initially very close to a fixed point solution
$u^*(y)$, i.e.
\eqn\ePerturb{u(y) = u^*(y) + v(y),}
where $v(y)$ is a small perturbation.  Inserting this into the flow
equation, Eq.\eFPfull, and keeping only terms first order in $v(y)$
gives
\eqn\eEVflow{\eqalign{
{{\del v} \over {\del t}} = & v + v'(1-{u^*}) + {u^*}'(v(0) - v) +
\beta y v' + \mu y(v'' + {u^*}''v(\!0\!)) \cr & + \mu\left[ 6y {u^*}'
v' + 2y^2({u^*}'v'' + {u^*}'' v') + y({u^*}'' (v - v(\!0\!)) +
({u^*}-1)v'' ) \right] ,\cr}} where we have chosen to perturb around
the fixed point with ${u^*}(0) = 1$.  Note that it is not permitted at
this point to choose $v(0)$, since it may not be a constant.  The
right hand side of Eq.\ePerturb\ can be thought of as a (nonlocal)
linear operator acting on $v(y)$.  Just as in finite-dimensional RGs,
solutions with simple exponential $t$ dependence can be
found (The question of completeness is discussed in the footnote [23].)
 if the spatially dependent part obeys the
eigenvalue equation:
\eqn\eEVstability{v + v'(1-u) + u'(1-v) + \beta y v' + \mu y(v'' +
u'') + \mu O(uv) = \lambda v,} where we have dropped the asterisk on
$u(y)$ and chosen $v(0) = 1$ : since eigenvectors are defined only up
to a constant, we have the freedom to choose a scale for $v(y)$. It is
straightforward to show that there are no eigenfunctions with $v(0) =
0$, the only choice not equivalent by a choice of scale to $v(0) = 1$.
This is shown in appendix B.  The terms in the square brackets in
Eq.\eEVflow\ are of order $\mu u v$.  For small $\mu$, they are
small for all $y$, and will be neglected in what follows.

As for the fixed point equation, the solutions to the eigenvalue
equation can be found in two regions and the pieces matched
asymptotically.  For $y \ll 1/\mu, 1/\beta$, the equation can be
rewritten in the standard form for first order linear differential
equations by dropping the $O(\mu uv)$ terms,
\eqn\eOuterstable{v'(y) + a(y)v(y) = b(y),}
where
\eqn\eABdef{\eqalign{
a(y) = & \; \; \left({{1 - \lambda - u'(y)} \over {1 - u(y)}}\right),
\cr
b(y) = & \; \; {{-u'(y)} \over {1 - u(y)}}. \cr}} Using the fact that
$u(y)$ satisfies Eq.\eNaive\ (with $\beta = 0$), and taking some care
due to the singularity in $u'(y)$ for small $y$, one finds the
solution
\eqn\eVouter{v = {1 \over \lambda}{{u^{1 - \lambda} - u} \over {1 -
u}},} for $y \ll 1/\mu, 1/\beta$.  In the matching regime, $y \gg 1$,
this solution behaves like
\eqn\eLeftmatchstab{v \sim e^{-1-y}\left[e^{\lambda y}
- 1\right]/\lambda .}

For $y \gg 1$, Eq.\eEVstability\ reduces to a linear equation very
similar to the fixed point equation (Eq.\eLinearized) in this regime,
\eqn\eLinstable{ \mu y v'' + (1 + \beta y)v' + (1 - \lambda)v = -u'
-\mu y u''.} As in the previous section (for the behavior as a
function of $\beta$), one expects a discrete set of exponential
solutions which match onto Eq.\eLeftmatchstab.  A natural substitution
is thus $v = e^{-y}w$, which yields \eqn\eWde{\mu y w'' + (1 -
\mu y)w' - \lambda w = (1 - \mu y)e^{-1},} using $\beta = \mu$ and $u =
e^{-1-y}$, which are valid for small $\mu$.  Expanding the solution in
a power series, $w(y) = \sum_m w_m y^m$, results in the following
recursion relations for the set $\{ w_m \}$: \eqn\eRecurse{\eqalign{
w_1 = & \; \; e^{-1} + \lambda w_0 , \cr w_2 = & {{-\mu e^{-1}} \over
{2(1+\mu)}} + {{\lambda + \mu} \over {2(1+\mu)}}w_1 , \cr w_{m+1} = &
{{\lambda + \mu m} \over {(m+1)(1 + \mu m)}}w_m \hskip 0.3in m \geq 2
. \cr}}

If the series does not terminate, then the recursion relation at high
order simplifies to $w_{m+1} \approx w_m/m$, so that $w_m
\sim 1/m!$.  This implies that for large $y$, $w(y) \sim e^{y}$
(times a power law of $y$ arising from the corrections to the
recurrence relation), so that the corresponding eigenfunction $v(y)$
has power-law decay.  This implies that the desired short-range
eigenfunctions form a discrete set, corresponding to the condition
that the series terminate at the $m^{\rm th}$ order, for $m =
1,2,\ldots$.

For $m \geq 2$, these conditions yield the eigenvalues $\lambda =
-n\mu$.  For $m=1$, the matching conditions fix $w_0$.  We guess that
$|\lambda| \ll 1$, and check for self consistency.  Under this
condition Eq.\eLeftmatchstab\ becomes $w \sim e^{-1}y$, so that $w_0 =
0$ and $w_1 = e^{-1}$, which satisfies the first recursion relation,
as it must for the solutions to match, and yields the condition
$\lambda = 0$ in order for $w_2$ to vanish.  This is thus just the
redundant marginal operator resulting from the choice of normalization
of the fixed point solution mentioned earlier.  The full set of
physical short-range eigenvalues is therefore
\eqn\eSRevalues{\lambda = -2\mu, -3\mu, -4\mu \ldots}
for small $\mu$.  Note that there is no eigenfunction with $\lambda =
-\mu$.  It is absent because of the inhomogeneous terms in Eq.\eWde,
as discussed above.

\newsec{Long-Range Correlated Disorder}

We now analyze the behavior for random potentials with long-range
power law correlations in $\bphi$.  We first consider the stability of
the short-range fixed point analyzed above to such long-range
correlated perturbations.

The behavior of the power-law eigenfunctions is much simpler than the
exponentially decaying solutions found above.  For any value of
$\lambda$ not in the discrete set of Eq.\eSRevalues, the solution
which is well behaved at the origin has power-law decay at infinity.
There is thus a continuum of power-law eigenfunctions with $v \sim
y^{-\Gamma}$, corresponding to $R(\bphi) \sim \phi^{-\gamma}$, with
$\Gamma = 1 + \gamma/2$.  The associated eigenvalues are
\eqn\eLRevalues{\lambda = 1 - \beta\Gamma,}
for all $\Gamma$ except those corresponding to the short-range
solutions (Eq.\eSRevalues), i.e. for
\eqn\eMissingevalues{\Gamma \neq {1 \over \beta},
{{(1-2\mu)} \over \beta}, {{(1-3\mu)} \over \beta}, {{(1-4\mu)} \over
\beta}, \ldots}
{}From Eq.\eLRevalues, we see that perturbations are
irrelevant for $\Gamma >1/\beta$.\ref\fCompleteness{Note that the
completeness of these eigenfunctions for the two classes of fixed
points remains to some extent an open question.  Since the linearized
operator in Eq.\eEVstability\ is not self-adjoint, standard completeness
theorems do not apply.  In general, one may look for different sets of
left and right eigenfunctions, and proceed in this way to examine
completeness and decompose arbitrary perturbations.  Unfortunately,
the non-locality of the operator implies that the associated adjoint
operator contains delta functions, and the analysis becomes
complicated by convergence issues. For the short-range fixed points,
however, completeness should hold, since the eigenfunctions are simply
exponentials multiplying polynomials, which should be equivalent to a
more standard basis by a simple orthogonalization.  For long-range
perturbations, much less can be stated.  It is unclear how to
decompose a particular perturbation in this basis, and also to what
extent short-range perturbations are representable in terms of such
functions.  We {\sl do} expect, however, that the behavior of the
eigenfunctions as a function of $\gamma$ does dictate the relevance of
perturbations with power-law tails.}

We now show that the behavior for distributions with long-range
correlations is, in fact, rather simple and general.  As seen in
section III, most solutions of the fixed point equation have power-law
tails.  It is only for certain special values of $\zeta$,
corresponding to the short-range eigenvalues, that the well-behaved
solution at $y=0$ connects to an exponentially decaying solution at
infinity.  In all other cases, the solution has a power-law decay for
large $y$, which is dictated entirely by the value of $\zeta$ and
$\epsilon$.

A simple large-$y$ analysis of Eq.\eFPfull\ gives the value of $\zeta$
quoted in the introduction
\eqn\eRLR{\zeta_{LR} = \epsilon/(4+\gamma),}
for
\eqn\eVLR{
\langle V(\bphi,\bx)V(\bphi',\bx') \rangle_C \sim |\bphi - \bphi'|^{-
\gamma}
\delta^{d}(\bx - \bx').}
It is easy to see that this result will be unaffected by the
higher-order terms in an RG expansion, since higher powers of
$R(\bphi)$ always appear with two derivatives, and multiplying
negative power-laws results in a more negative power-law.  (We restrict
our attention to $\gamma > -2$, which is needed to make the problem
well defined.)  Thus, at least within the perturbative RG, only the
scale change terms in the RG flows (i.e. those in Eq.\eFlowR\ that are
multiplied by $\epsilon$ and $\zeta$) are needed to fix $\zeta$.  The
exponent $\gamma$ thus fixes the roughness exponent at the long-range
fixed point exactly.  We believe that this result should be strictly
true in all dimensions, but an actual non-perturbative proof would
clearly be desirable.

For Eq.\eLRevalues, we see that the condition that the long-range
correlations dominate and that Eq.\eRLR\ applies is that $\gamma <
1/\beta_{SR}$, implying $\gamma < \gamma_c \equiv \epsilon/\zeta_{SR}
- 4$.  Since this also arises from just the rescaling part of the RG
flows, we expect it to be true in all dimensions less than four.  As
$\gamma$ decreases, the short-range fixed point will become unstable
when $\gamma = \gamma_c$; and for $\gamma < \gamma_c$, $\zeta$ will
vary continuously away from $\zeta_{SR}$ according to Eq.\eRLR.

A special case of long-range correlated randomness corresponds to
interfaces in random field systems which have $N=1$ and $\gamma = -1$.
The general result of Eq.\eRLR\ implies that $\zeta_{RF} = (4-d)/3$ as
obtained by many authors.\cite{\rIM-\rDSFFRG}

Since the short-range fixed point becomes linearly unstable for
$\gamma$ below $\gamma_c$, it should be possible to observe the
instability of the long-range fixed point in the opposite regime, as
$\gamma$ increases to $\gamma_c$.  This is actually a subtle problem.
If one proceeds with a naive calculation of the eigenfunctions around
the long-range fixed point, one arrives again at Eq.\eEVstability.  In
this case, however, a simple argument demonstrates that {\sl all} the
solutions have power-law form.  For large $y$, Eq.\eEVstability\
becomes
\eqn\ePowerlawlargey{\mu y v'' + (1 + \beta y)v' + (1 - \lambda)v \approx
A\Gamma \left[1 -\mu(\Gamma+1)\right]y^{-\Gamma-1},} using $u^*(y)
\sim Ay^{-\Gamma}$ for large $y$.  Since the power-law dictated by the
fixed-point function appears on the right-hand side, the homogeneous
terms must balance this, and the only way they can do so is to develop
power-law tails themselves.  Therefore, all the eigenfunctions have
power-law decay at large $y$.  This means that the growth of
short-range correlations does {\sl not} manifest itself in the usual
manner as a relevant eigenvalue.  Although we have not calculated this
explicitly within the framework of the large $N$ RG, we expect the
following behavior: as $\gamma$ increases towards $\gamma_c$, the
fixed point correlation function of the random potential will become
more and more like the short-range fixed point with the regime in
which the power law tail appears moving out to larger and larger $y$,
eventually disappearing for $\gamma >\gamma_c$.\ref\fHHcritique{It is the
non-uniformity of this limit which led to the incorrect results of
ref.\cite{\rHH}.  There, it was assumed that the growth of short-range
correlations occurs via the appearance of an exponential damping of
the long-range power law, implying that the power-law prefactor appearing in
Eq.\eLargey\ is equal to $y^{-\gamma_c}$.  A comparison of the large
$N$ results of Eq.\eRoughfinalSR\ and Eq.\eLargey\ demonstrates
explicitly the incorrectness of this assumption.}

\newsec{Conclusions}

In this paper we have studied the problem of an oriented manifold with
$d$ internal and $N$ transverse dimensions.  For $d$ near 4, it became
possible to treat the zero temperature fixed point in an expansion in
$\epsilon = 4-d$, in terms of a second-order non-linear differential
equation.  At $N=\infty$, this equation could be solved exactly for
the most interesting case of short-range disorder, yielding a
roughness exponent $\zeta_{SR} = \epsilon/(4+N)$.  For large but
finite $N$, boundary-layer techniques were employed to estimate the
leading corrections, which were found to be non-analytic, vanishing as
$2^{-N}$.  The magnitude of this non-analytic correction for $N=1$ is
comparable to the correction to $\epsilon/5$ found numerically: $\zeta
= 0.2083\epsilon$.  The behavior in large $N$ near four dimensions
saturates a lower bound $\zeta \geq (4-d)/(4+N)$ that is believed to
be exact.\ref\rDSFunpub{D. S. Fisher, unpublished.}\ The leading
correction to this that we have found is indeed positive as it should
be.  Formally, preliminary analysis of our RG flow equations in the
opposite limit yields $\zeta(N=0) =
\epsilon/4$.  This is believed to be an exact upper bound for general
$N$\cite{\rDSFunpub}.  All known numerical and analytical results do indeed lie
in
the range
\eqn\eZetarange{ {{4-d} \over {4+N}} \leq \zeta \leq {{4-d} \over 4}.}

It is interesting that our large $N$ $\epsilon$-expansion result
agrees with the large $N$ results of M\'ezard and Parisi, claimed to
be valid for general dimension ($2<d<4$) by replica symmetry breaking
techniques. A preliminary large $N$ analysis of the zero temperature
minimization problem corresponding to our RG procedure does {\sl not}
suggest that the $O(\epsilon)$ result should be exact for $\epsilon
<2$ in this limit, although further study may change this conclusion.
The analysis does, however, suggest that the next corrections to
$\zeta$, which are probably of $O(\epsilon^{3/2})$, may be calculable
without taking into account the effects of multiple local minima.

Both extensions of the $\epsilon$-expansion and investigation of the
large $N$ limit beyond the $\epsilon$-expansion are worthwhile future
endeavors.  An important question is whether or not the
``replica-symmetry breaking'' used in the variational ansatz by Mezard
and Parisi has any well-defined physical interpretation beyond that of
the general scaling picture of manifolds in random media (discussed in
detail in the directed polymer context by Fisher and Huse\cite{\rFH}).
Answering this might bear fruit for understanding other random
systems, such as spin and vortex glasses.  Finally, it is possible
that some of the techniques used here might be applicable for other
problems such as periodic (e.g. charge density wave or flux lattices)
or non-periodic (e.g. polymerized membranes) elastic manifolds which
exist in random media of the same dimension.

\centerline {\bf Acknowledgments}
This work is supported in part by the National Science Foundation,
through Harvard University's Materials Research Lab, grant DMR9106237
and a graduate fellowship to L.B..  D. S. F. was also supported by the
A. P. Sloan Foundation.

\appendix{A}{Perturbative Calculation in Primary Region}

For the primary region, $y \ll 1/\mu$, simple perturbation theory can
be used to find the effects of the $O(\mu)$ terms of Eq.\eFPfull.  The
solution is expanded in a power series in $\mu$,
\eqn\eMuexpand{u = u_\infty + \mu u_1 + \mu^2 u_2 + \ldots,}
and terms are grouped order by order, assuming (as confirmed in
section III) that $\beta = O(\mu)$.  The zeroth order equation is just
Eq.\eNaive, while the first order terms give
\eqn\eFOpert{u'_1 (1 - u_\infty) + u_1 (1 - u'_\infty) = 3y(u'_\infty)^2 + 2y^2
u'_\infty u''_\infty + y u''_\infty(u_\infty -1) - (\beta/\mu) y
u'_\infty,} where $u_\infty(y)$ is the solution of Eq.\eNaive.  The
right hand side of this equation can be rewritten completely in terms
of $u_\infty$ by using Eq.\eImplicit\ to eliminate the $y$ dependence,
and Eq.\eNaive\ to eliminate derivative terms.  After some lengthy
algebra, one finds
\eqn\eRHSsimpler{u'_1 (1 - u_\infty) + u_1 (1 - u'_\infty) = H(u_\infty),}
where
\eqn\eHdef{
H(x) \equiv H_0(x) + H_\beta (x), } with
\eqn\eHsubdef{\eqalign{
H_0(x+1) = & \left[ - 2/x^2 -2/x + 3x + 3 \right] + \log (x+1) \left[
4/x^3 + 6/x^2 -1/x -3 \right] \cr & + \log^2(x+1) \left[ -2/x^4 -
4/x^3 -2/x^2 \right], \cr {\rm and} \quad H_\beta(x) = & -{\beta \over
\mu} \left[ x - \log(x) - \log(x)/(x-1)
\right].\cr}}
Since Eq.\eRHSsimpler\ now has no explicit $y$ dependence, we switch
to the dependent variable $u_\infty$, using
\eqn\eVarswitch{ u'_1 \equiv {du_1 \over {dy}} = {du_1 \over
{du_\infty}}u'_\infty
= {du_1 \over {du_\infty}} {u_\infty \over {u_\infty -1}}.}
Eq.\eRHSsimpler\ becomes
\eqn\ePTsimpler{ {{du_1} \over {du_\infty}} + {u_1 \over
{u_\infty(u_\infty-1)}} =
-H(u_\infty)/u_\infty,} which has the solution
\eqn\eUonesolved{u_1(u_\infty) = {u_\infty \over {1-u_\infty}}\int_{u_\infty}^1
{{1-x}
\over x^2}
H(x) dx,} where the boundary condition $u_1(y=0) = 0$ has been
imposed.  The important limit for the matching carried out in section
3 is $y \gg 1$, corresponding to $u_\infty \sim \exp(-1-y) \goes 0$.
In this limit, the integral yields
\eqn\eUonesolution{ \mu u_1(y) \approx e^{-1-y} \left[ {1 \over 2}(\beta -
\mu)(y^2 -1) - \mu \right].}
In section III, it was found that $|\beta - \mu| \ll \mu$, so that the
first term in the brackets can be neglected.  The full solution is
then still a pure exponential in the matching region, but with a
different coefficient:
\eqn\eFullmatchingsolution{u(y) \approx e^{-1-y}(1 - \mu) \hskip 0.2 in
{\rm for} \hskip 0.1 in 1 \ll y \ll 1/\mu.}

\appendix{B}{Eigenfunctions with $v(0) = 0$}

By choosing $v(0) = 0$, we arrive at an equation for this case
analogous to Eq.\eEVstability,
\eqn\eEVvzero{v + (1-u)v' - u' v + \beta y v' + \mu y (v'' + u'') = \lambda
v,} where we have already neglected the $\mu O(uv)$ terms.  In the
perturbative region ($y \ll 1/\mu$), the equation analogous to
Eq.\eOuterstable\ is
\eqn\eOutvzero{v'(1-u) + (1-\lambda -u')v = 0,}
which in this case is homogeneous, and correspondingly simpler to
solve.  Using Eq.\eNaive and Eq.\eImplicit, one finds the general
solution
\eqn\eOutvzerosolved{v = Cu^{1-\lambda}/(1-u).}
As $y \goes 0$, $u \goes 1 - \sqrt{2y}$, so that
\eqn\eVzerosmall{ v(y) \goes {C \over \sqrt{2y}}.}
For $v$ to be non-singular for small $y$, the only choice is $C=0$, so
that no such eigenfunctions exist.

This is in some respects a surprising result, since for any
perturbation of $u^*(y)$, one should be able to rescale the resulting
function to leave $u(0)$ unchanged.  Therefore one might expect the
stability analysis to be phrased precisely in terms of those
perturbations which {\sl do not} change $u(0)$.  The rescaling,
however, is equivalent to adding some amount of the marginal
eigenfunction which moves along the fixed line.  The function
representing the combined effects of perturbation and rescaling (with
$v(0) = 0$), is, in this case, not an eigenfunction.

\appendix{C}{Schematic Minimization}

To understand the effects of multiple minima in the random potential
on the validity of the RG used here, it is instructive to consider a
simple model problem representing the iterative minimization at a
particular length scale.  This provides a physical derivation of the
renormalization group flow equation (Eq.\eRGR).  Schematically, we
imagine integrating out a single Fourier mode $\bphi_>$ of momentum
${\bf p}$.  At zero temperature, this reduces to the problem of
minimization over this $N-$dimensional vector.  The renormalized
potential will then be given by
\eqn\eMinproblem{V_R(\bphi_<) = \min_{\bphi_>} \left\{ {1 \over 2}
\phi_>^2 +
V(\bphi_< + \bphi_>) \right\},} where the magnitude of the momentum
cut-off $p = \Lambda$ has been set to one for convenience, and, more
importantly, we have ignored the spatial dependence of the potential
$V(\bphi,\bx)$.  A similar formulation of the RG would appear in the
treatment of elastic manifolds on a hierarchical lattice.  Although
Eq.\eMinproblem\ may appear to be an unreasonable approximation, we
will find that the lowest RG order flow equations that we have used in
this paper are exactly reproduced.  Indeed, the treatment of a
suitably modified version of Eq.\eMinproblem\ which takes into account
the spatial dependence does not differ substantially from the
approximate version described here (though the required notation makes
it rather more cumbersome).

For the purposes of simplicity, we will concentrate on the case
of $N=1$, which, for small $\epsilon$, does not differ much from the
case of general $N$.  For the remaining part of this appendix, we
introduce the notation $x \equiv \phi_<$ and $y
\equiv \phi_>$.  The $N=1$ problem is then
\eqn\eNonemin{\epsilon^{1/2}V_R(x) = \min_y U(x,y) \equiv \min_{y}
\left\{ {1
\over 2} y^2 +  \epsilon^{1/2}V(x+y)
\right\},}
where a factor of $\epsilon^{1/2}$ has been extracted to make $V$ of
order $1$.  To estimate quantities, we use the fixed-point values for
the correlation function of $V(y)$,
\eqn\eVcorrelation{\eqalign{
\langle V(x)V(x') \rangle = & \tilde{R}(x-x') \equiv R(x-x')/\epsilon, \cr
\langle V(x) \rangle = & 0.\cr}}
Because the minima in Eq.\eNonemin\ will be at small $y$ for small
$\epsilon$, it is the small-distance behavior of $\tilde{R}(x)$ which
will determine the behavior of the model.  Based on the behavior of
the fixed point function $R(x)$ in the RG of section II, we assume that
$\tilde{R}(x)$ can have a discontinuity in its third derivative at
$x=0$.

The extremal condition for Eq.\eNonemin\ is
\eqn\eExtremalcond{ y = \epsilon^{1/2} F(x+y),}
where
\eqn\eForcedef{ F(y) \equiv -V'(y)}
is the force at the ``position'' $y$, and primes have been introduced
to denote derivatives.  From Eq.\eVcorrelation\ the correlations of
the force at short distances are \eqn\eForcecorrelations{\eqalign{
\left\langle F(x)^2 \right\rangle = & 1, \cr
\left\langle (F(x) - F(0))^2 \right\rangle \sim & |x|, \cr}}
for $|x| \ll 1$.

The perturbative RG performed in section II is equivalent to assuming
a perturbation series for $y$ in $\epsilon^{1/2}$ in
Eq.\eExtremalcond\ and expanding $F(x+y)$ to obtain solutions order by
order in $\epsilon^{1/2}$.  There are two possible ways in which such
an expansion may break down.  Most obviously, the linear behavior in
Eq.\eForcecorrelations\ implies that derivatives of $F(y)$ are
typically infinite, so that the analyticity assumed by a power series
in $\epsilon^{1/2}$ may break down.  Secondly, the extremal condition
(Eq.\eExtremalcond) may have multiple solutions, and an iterative
solution may converge to any of these, including both maxima and
non-global minima.

Note that if $\tilde{R}(x)$ were smooth, then $\langle [V''(x)]^2
\rangle < \infty$ and the full potential $U$ in Eq.\eNonemin\ would
have strictly positive curvature with high probability for small
$\epsilon$ and hence a unique minimum.  The apparent assumption of
analyticity of $\tilde{R}(x)$ may be circumvented by the use of a
different iterative procedure to find the minimum of the potential
$U$.  The simplest such method, which reproduces the results of the
perturbation series for the analytic case, is a version of gradient
descent (see Fig.[3]) One defines a sequence of approximants $\{y_0,
y_1, y_2, \ldots \}$ to Eq.\eExtremalcond\ by
\eqn\eIterativeapprox{y_{n+1} = \epsilon^{1/2}F(x + y_n),} for
$n=0,1,2,\ldots$

It is interesting to note that the stability properties of such a
mapping discriminate between minima and maxima.  Letting $y_n = y^* +
\delta y$ and linearizing, one finds
\eqn\eLinearizedmapping{\delta y_{n+1} = -V''(y^*)\delta y_n,}
which is stable for $|V''(y^*)| < 1$.  Since the curvature of the
potential $U$ in Eq.\eNonemin\ is just $1+V''(y)$, this condition
excludes all maxima (as well as minima which are sufficiently narrow).
Note, however, that the iterative scheme does not guarantee
convergence, and one cannot rule out limit cycles or other more
complicated behavior for particular realizations of the disorder.
Furthermore, even if it does converge, it may not be to the desired
global minimum.  As we shall see, however, we will obtain an estimate
of the global minimum with sufficient accuracy for our present
purposes.

Iterating Eq.\eIterativeapprox\ yields the first few approximants,
\eqn\eFirstfewapproxs{\eqalign{
y_0 = & 0 , \cr y_1 = & \epsilon^{1/2}F(x) , \cr y_2 = &
\epsilon^{1/2}F(x + \epsilon^{1/2}F(x)), \cr {\rm etc} & \cr}} Using
Eq.\eForcecorrelations\ the corrections at each level of approximation
may be estimated.  From Eq.\eIterativeapprox,
\eqn\eCorrectionestimate{\eqalign{
y_{n+1} - y_n = & \epsilon^{1/2}\left[ F(x+y_n) - F(x+y_{n-1})\right]
\cr \sim & \epsilon^{1/2}|y_n - y_{n-1}|^{1/2},\cr}} with a random
coefficient of $O(1)$.  Iterating this, an infinite series of
non-trivial powers of $\epsilon$ appear.  For the $n^{\rm th}$
approximant, we thus have
\eqn\eNthapproxestimate{y_n \approx A\epsilon^{1/2} + B\epsilon^{3/4}
+C\epsilon^{7/8} + \ldots + Z\epsilon^{(2^n - 1)/2^n}.} The difference
from the perturbation series result for the analytic case appears
first in $y_2$, via the appearance of the $\epsilon^{3/4}$ term in
Eq.\eNthapproxestimate.

{}From Eq.\eExtremalcond, we may find an upper bound for the size of the
region in which minima are likely to exist.  (In fact, we only
calculate this bound for the region within which there are {\sl
extrema}.  For $N=1$, the furthest out extrema are in fact always
minima, but for large $N$ this difference may be important.)  Suppose
one extrema is located at the point $y^*$, satisfying
Eq.\eExtremalcond.  Then for another extremum to be located within a
distance $\delta y$, the condition must again be satisfied at the
point $y+\delta y$.  For large $\delta y$, this is clearly extremely
unlikely, since the linear term grows while the random force remains
bounded and of $O(\epsilon^{1/2})$ with high probability.  For small
$\delta y$, the variations of the force grow like $(\delta y)^{1/2}$
from Eq.\eForcecorrelations, i.e.  {\sl faster} than the linear term
in Eq.\eExtremalcond.  Thus there will be a length scale below which
the variations of the force dominate, and other extrema are possible.
Equating the two terms in Eq.\eExtremalcond\ gives
\eqn\eExtremabound{
\delta y \sim \epsilon^{1/2}|\delta y|^{1/2}  \Rightarrow
 \delta y < O(\epsilon), } which is the desired upper bound on the
separation of extrema.  We see that the separation between extrema is
smaller than {\sl any} of the correction terms obtained in
\eNthapproxestimate. This is illustrated in Fig.[3]. This suggests that the
corrections due to
multiple minima appear at higher order in $\epsilon$ than the
iterative corrections which arise from the non-analyticity of $V(y)$.
To check this, we must analyze how the corrections to $y$ affect the
renormalized potential and its correlations.

To estimate the corrections from the terms in Eq.\eNthapproxestimate,
it is useful to write the random potential is the form
\eqn\ePotentialrewritten{V(x+y) =
V(x) + yV'(x) + W(x;y),} where
\eqn\eWdef{W(x;y) \equiv \int_0^y \left[ V'(x+z)  - V'(x) \right] dz.}
The renormalized potential can then be evaluated by inserting the
iterative solution Eq.\eNthapproxestimate\ into Eq.\eNonemin, yielding
\eqn\eVrenormalized{\eqalign{\epsilon^{1/2}V_R(\! x \! ) = & \epsilon^{1/2}
V(\! x \! ) + {{A^2(\! x \! )} \over 2}\epsilon + {{B^2(\! x \! )}
\over 2}\epsilon^{3/2} + {{C^2(\! x \! )} \over 2}\epsilon^{7/4} +
\cdots \cr &+ A(\! x \! )B(\! x \!  )\epsilon^{5/4} + A(\! x \! )C(\!
x \! )\epsilon^{11/8} + \cdots + \epsilon^{1/2}V'(\! x \! )\left[ A(\!
x \!  )\epsilon^{1/2} + B(\! x \! )\epsilon^{3/4} + \cdots \right] \cr
& + \epsilon^{1/2}W[x; A(\! x \! )\epsilon^{1/2} + B(\! x \!
)\epsilon^{3/4} +
\cdots ].\cr}}

This expression simplifies somewhat when the function $A(x) = F(x) = -V'(x)$
from Eq.\eFirstfewapproxs\ is inserted,
\eqn\eVrenormsimpler{\eqalign{
\epsilon^{1/2}V_R(\! x \!) = & \epsilon^{1/2} V(\! x \!) - {{A^2(\! x \! )}
\over 2}\epsilon +
{{B^2(\! x \! )} \over 2}\epsilon^{3/2} + \cdots \cr & + B(\! x \!)C(\! x
\!)\epsilon^{13/8} + \cdots
 + \epsilon^{1/2}W[ A(\! x \!  )\epsilon^{1/2} + B(\! x \!
)\epsilon^{3/4} + \cdots ]. \cr}}
The fact that the second term is negative reflects the approach to the
minimum.  Note that the cross terms $AC, AD, etc.$ have now
canceled.  The RG flow requires renormalization of the
correlation function of the disorder, Eq.\eVcorrelation.  The
renormalized potential in Eq.\eVrenormsimpler\ will have a non-zero
expectation value, since the minimization procedure decreases the
energy for all realizations of $V(y)$.  To find the renormalized
correlation function, therefore, it is necessary to take the truncated
(cumulant) expectation value,
\eqn\eRrenormalizedcum{\eqalign{
\epsilon R_R(x)\! = & \epsilon\langle V_R(x) V_R(0) \rangle_C \cr
= & \epsilon \langle V(x) V(0) \rangle\! -\!{1 \over
2}\epsilon^{3/2}\langle A^2(x)V(0)\!+\!A^2(0)V(x)\rangle_C\! +\! {1 \over
2}\epsilon^2 \langle B^2(x)V(0) + B^2(0)V(x)\rangle_C \cr
& \!+\! {1 \over 4} \epsilon^2 \!\langle
A^2(x)A^2(0) \rangle_C \!+\! 2\epsilon\langle V(x) \hat{W}[0]
\rangle_C \epsilon \!+\! \epsilon\langle \hat{W}[x]\hat{W}[0] \rangle_C \cr
& \!-\! {1 \over 2}\epsilon^{3/2}\langle A^2(x)\hat{W}[0] \!+\!
A^2(0)\hat{W}[x]\rangle_C \!+\! {1 \over 2}\epsilon^2 \langle
B^2(x)\hat{W}[0] \!+\! B^2(0)\hat{W}[x] \rangle_C \!+\!
O(\!\epsilon^{17/8}),\cr
}}
 where in this expression $\hat{W}[x]$ means the full expression from the
last term of Eq.\eVrenormsimpler, evaluating all internal coefficients
(i.e.  $A(x), B(x)$ etc.) at the point $x$, and we have dropped all
terms of explicitly higher order than $\epsilon^2$.  The function
$\hat{W}[x]$ is, however, itself small, so that some further terms can
be dropped.  From Using Eq.\eWdef\ and the fact that $F(x) - F(x+z) \sim
\sqrt{z}$ (Eq.\eForcecorrelations), we see that $W(x;y) \sim
|y|^{3/2}$.  Since $y
\sim \epsilon^{1/2}$, $\hat{W}[x] \sim \epsilon^{3/4} + O(\epsilon)$,
where the $O(\epsilon)$ term arises from the $B(x)$ in
Eq.\eNthapproxestimate\ and higher order terms in $A(x)$; this
$O(\epsilon)$ term is needed to obtain the $V_R$ correlations to
$O(\epsilon^2)$ but higher order terms are not.  In addition we see
that of all the terms involving $\hat{W}$ in Eq.\eRrenormalizedcum,
only the $V\hat{W}$ term will contribute at this order.  We thus see
that to $O(\epsilon^2)$, we may drop all terms in $y$ beyond the
$B(x)$ term.  To this order, we may thus use the second order
iterative solution
\eqn\eSOIS{\epsilon^{1/2}V_R(x) \approx
\epsilon^{1/2}V\left[x+\epsilon^{1/2}F[x + \epsilon^{1/2}F(x)]\right]
+ {1 \over 2}\epsilon\left[ F[x+\epsilon^{1/2}F(x)] \right]^2.}
The correlations of $V_R(x)$ and $V_R(0)$ can be calculated directly
from this form of $V_R$ yielding
\eqn\eRrenormdirect{R_R(x) = \epsilon\langle V_R(x)V_R(0) \rangle_C =
R(x) + {1 \over 2}\left[R''(x)\right]^2 - R''(x)R''(x) +
O(\epsilon^{5/2}).}
To this order, the correct answer can be obtained from
Eq.\eRrenormalizedcum\ by expanding $W(x;y)$ formally as $W(x;y)
\approx {y^2 \over 2}V''(x)$ and averaging directly the $\hat{W}V$ term in
Eq.\eRrenormalizedcum.  This is valid because only one $V''$ appears
here.  To analyze the $\hat{W}\hat{W}$ term, an expansion in $y$ {\sl
fails} and $\langle \hat{W}\hat{W} \rangle \sim \epsilon^{3/2}$ rather
than the naive $\epsilon^2$.  We thus expect that the effects of the
nonanalyticity of $R$ will affect $R_R$ at order $\epsilon^{5/2}$.
These terms need to be balanced by adjustments to $\zeta$, suggesting
$O(\epsilon^{3/2})$ corrections to our $O(\epsilon)$ result for $\zeta$.

So far, the effects of multiple extrema have not been included.  Their
effects can be estimated by including a further correction term in
$y$,
\eqn\eMultiplecorrectionterm{y = A\epsilon^{1/2} + B\epsilon^{3/4} +
\cdots + \eta\epsilon + \cdots}
By repeating arguments along the lines of those above, it is a simple
matter to estimate the leading corrections due to a non-zero $\eta$.
One finds that the first contribution to $R(x)$ occurs at
$O(\epsilon^3)$.  This is higher order than all the leading
corrections from the non-analyticities arising in the iterative procedure.

Thus, although the singularity in $R'''(x)$ at $x=0$ is associated
with the existence of many extrema (since $V''(0)$ has infinite
variance), the direct effects of these multiple extrema only show up
at higher order in $\epsilon$;  to the order needed here, choosing
{\sl any} of the minima provides enough accuracy.

The simple approximation to the renormalization group flows analyzed
in this Appendix suggests that there will be corrections
to $\zeta$, starting at $O(\epsilon^{3/2})$, with an apparently
infinite sequence of higher order corrections appearing before
$O(\epsilon^2)$, at which order the effects of multiple minima begin to
appear.  Although the picture is quite appealing, the results should
not be taken as definitive predictions of the powers involved, since a
complete analysis should involve a self-consistency condition to
determine the small $x$ behavior of the correlations.  Such an
analysis may well involve a boundary layer for small $x$ with
smoothing of the fixed-point function on scales smaller than
some ($>1$) power of $\epsilon$.

It is straightforward to extend the analysis of this Appendix to
general (fixed) $N$ in the limit of small $\epsilon$.  Since in the
limit of large $N$, $\zeta$ is formally small, even if $\epsilon$ is
not small, one might hope to be able to justify truncation of the RG
flows for all $\epsilon$ (or at least $\epsilon <2$) for $N$ large.  We
have not been able to do this, and, indeed, preliminary indications
suggest the opposite conclusion: that even for large $N$ higher order
terms in $\epsilon$ are needed.  A more detailed study of this limit
would clearly be instructive.

\appendix{D}{Multicritical Short-Range Fixed Points}

As remarked in section III, Eq.\eLinearized\ possesses a discrete
family of solutions which are well-behaved at the origin and decay
exponentially at infinity.  These can be matched onto the primary
solution to yield additional fixed points of the RG flows. To find
these, we perform a power series expansion similar to the one used for
the stability analysis (Eq.\eRecurse).  Defining
\eqn\eWdef{\eqalign{
u \equiv & \; \; e^{-{\beta \over \mu} y} w, \cr w = & \; \; \sum_m
w_m y^m, \cr}} Eq.\eLinearized\ yields a simple recursion relation for
the set $\{ w_m \}$:
\eqn\eWrecurse{ w_{m+1} = \left[{ {\beta/\mu - 1 + \beta m} \over {
(m+1)(1 + \mu m)}} \right] w_m.} If the series does not terminate, the
large $m$ behavior of the coefficients is
\eqn\eLargencoeff{w_m\sim {1 \over {m!}}\left({\beta \over \mu}
\right)^m,}
so that $u(y)$ decays more slowly than an exponential.  A short-ranged
$u(y)$ is obtained whenever the series terminates, which yields the
condition
\eqn\eBetacond{\beta = {\mu \over {1 + m\mu}} \hskip 0.3 in
m=0,1,2,\ldots.} In terms of the roughening exponent,
\eqn\eZetacond{\zeta = {\epsilon \over {4 + N + 2m}}.}
The case $m=0$ corresponds to the simple exponential found in section
III, while for higher $m$ the solutions have some oscillations and
correspond to smaller values of
$\zeta$.

It is a simple matter to extend the results of section IV to calculate
the stability around the new fixed points.  One finds that for the
$m^{th}$ fixed point, there are $m$ relevant eigenvalues corresponding
to short-range correlated perturbations, so that the
solution found in section III is stable, while the remaining solutions
represent a hierarchy of multicritical solutions.

Although such solutions exist formally, we have not fully investigated
the criteria under which these solutions represent physically
meaningful fixed points.  At least initially, the function $R(\bphi)$
is highly constrained by the positivity condition for the probability
distribution of $V(\bphi)$.  In particular, taking the Fourier
transform of the potential-potential correlation function, we must have
\eqn\eVVtransform{\left\langle \tilde{V}({\bk})\tilde{V}({-\bk})
\right\rangle = \tilde{R}({\bk}) \ge 0.}
Although the interpretation as a correlation function suggests that
this positivity property is preserved by the RG, the non-locality (in
$\bk$) of the terms in the RG flows generated by fluctuations has
prevented us from finding a simple proof.  Nevertheless, it seems
likely that the positivity is preserved.  It is straightforward to
check the multicritical fixed points obtained above for this
criterion.  If they do not satisfy Eq.\eVVtransform\ they cannot be
physical.  A simple computation for the first multicritical solution,
\eqn\eFirstmulticriticalsolution{R_1(\bphi) = \left(1 - \mu^2 - {{\mu \phi^2}
\over 2} \right) \exp \left( - {\phi^2 \over {2(1+\mu)}} \right),}
yields the Fourier transform
\eqn\eFirstmultsolFT{\tilde{R}_1({\bk}) = \left( {1 \over 2} + {\mu
\over 2} (1 + \mu)^2\kappa^2 \right) \exp \left( - {\kappa^2 \over
2}(1+\mu) \right),}
in the large $N$ limit, which satisfies the positivity criterion
(Eq.\eVVtransform) and thus might be physically attainable.

If we restrict consideration to distributions in which the function
$u(y)$ has no zeros (or equivalently $R(|\bphi|)$ has no non-trivial
extrema), it is possible to show, however, that the multi-critical
solutions, which have at least one zero are inaccessible.  A simple
argument proceeds as follows (see Fig.[4]): Consider the evolution of
an initial function $u_0(y)$ which has no zero crossings.  For the
function $u_0(y)$ to evolve into one of these multicritical solutions,
it must at some intermediate stage when it first has a zero be tangent
with the $y$ axis at some point $y_i$ (like the function $u_i(y)$ in
Fig.[4]).  (Note that since the behavior at $\infty$ is preserved by
the flows, the zero cannot come in from $\infty$ and
avoid the tangency condition.)  At this intermediate point, the
function must obey
\eqn\eZerocrossingintermediatept{\eqalign{
u_i(y_i;t_i) = & 0, \cr u'_i(y_i;t_i) = & 0, \cr
{\rm and} \; u''_i(y_i,t_i) > & 0. \cr}}
{}From Eq.\eFPfull, one then finds
\eqn\eZerocrossingtimechange{\left.{{\del u(y_i;t)} \over {\del t}}\right|
_{t = t_i} = \mu u_0(t_i)y_i u''(y_i,t_i) > 0,}
so that the putative point disappears.  Therefore such a zero cannot occur,
and any function $u(y)$ with zeros is inaccessible from an initial $u$
without zeros.

At this point it is unclear whether the formal multicritical fixed
points found here are accessible for less restrictive initial
correlation functions, and, if so, what is their physical
significance.  In particular, one might expect the $m=1$ critical
point to separate two phases with different behavior.  If one of these
is the rough phase analyzed in this paper, what is the nature of the
other phase?  We leave these as intriguing open questions.

\vfill\eject\immediate\closeout\rfile                       %
\centerline{{\bf References}}\bigskip                                     %
{\catcode`\@=11\escapechar=` \input refs.tmp\vfill\eject}

\figures
\fig{1}{Diagrams representing the generation and feedback of the
three-replica term $S$ (see the third line of Eq.\ePCflow).  Fig.[1a]
shows how such an operator is generated at second order in $R$ by
terms with one contraction.  Since the momentum of the internal line
is within the momentum shell, the diagram only contributes at large
momentum.  Summing up all such terms resulting from the expansion of
$R$ (Eq.\eRpowerseries) gives Eq.\eThreereplica.  Since the
fixed-point value of $S$ is $O(\epsilon^2)$, the only potentially
dangerous contribution for our analysis is first order in $S$.  This
feedback comes from diagrams such as that of Fig.[1b], with a single
loop.  Since the original $S$ vertices were generated only at high
momentum, such terms do not renormalize $R$. }

\fig{2}{Regions of validity of the primary (perturbative) solution,
Eq.\eNaive, and the tail (linearized) solution, Eq.\eLinearized.  In
the large $N$ (small $\mu$) limit, the size of the matching region ($1
\ll y \ll 1/\mu$) grows without bound.}

\fig{3}{Graphical illustration of the iterative minimization of
Eq.\eIterativeapprox.  Given a guess $y_i$ for the location of the
minima, the next approximation is found by following a vertical line
at this value of $y$ until it intersects the random force curve.
Extending a horizontal line to the $45^\circ$ line through the origin
(representing the uniform restoring force of the harmonic potential)
gives the value of $y$ for the next iteration.  The second two
approximants, $y_1$ and $y_2$, resulting from the initial $y_0=0$ are
shown here.  The many intersections between the random force curve and
the $45^\circ$ line in the figure represent multiple extrema which
occur on smaller scales (see Eq.\eMultiplecorrectionterm).}

\fig{4}{Illustration of the preservation of the lack of zeros by the RG flows
(Eq.\eFPfull) for the function $u(y)$.  For the initial function
$u_0(y)$ to develop into the final function $u_f(y)$, with internal
zeros, it must pass through an intermediate stage $u_i(y)$, at which
it is tangent at some point $y_i$ with the $y$ axis.
Eq.\eZerocrossingtimechange\ shows that such a point of tangency is
repelled, so that the putative crossing does not occur.} \end